\documentstyle[aps,12pt]{revtex}

\begin{document}
\draft
\author{Sergio De Filippo}
\author{Filippo Maimone}
\address{Dipartimento di Fisica ''E.R. Caianiello'', Universit\`{a} di Salerno\\
Via Allende I-84081 Baronissi (SA) ITALY\\
Tel: +39 089 965 229, Fax: +39 089 965 275, e-mail: defilippo@sa.infn.it\\
and \\
Unit\`{a} I.N.F.M., I.N.F.N. Salerno}
\date{\today}
\title{A class of nonunitary models of Newtonian gravity and its unicity. }
\maketitle

\begin{abstract}
A class of non-Markoffian nonunitary models for Newtonian gravity is
characterized as following from some rather natural hypotheses. One of such
models was previously obtained as the Newtonian limit of a classically
stable version of higher derivative gravity. They give rise to a mass
threshold around $10^{11}$ proton masses for gravity induced localization,
to a breaking of linearity and to the possible identification of
thermodynamic and von Neumann entropies.
\end{abstract}


In a recent paper a nonlinear nonunitary model of Newtonian gravity was
obtained as the nonrelativistic limit of a classically stable version of
higher derivative gravity\cite{DefMaim}. While reproducing at macroscopic
level the ordinary Newtonian interaction, it presents a mass threshold for
gravitational localization, which for ordinary matter densities is around $%
10^{11}$ proton masses. This model can be seen as a realistic version of the
nonunitary toy models\cite{ellis,banks,unruh} inspired by the emergence of
the information loss paradox\cite{hawking1,preskill,wald} from black hole
physics. On the other hand the violation of unitarity when matching quantum
mechanics and gravity was argued even outside black hole physics, on general
consistency grounds\cite{karolyhazy,penrose}. In fact the existence of
linear superpositions of states with macroscopic mass-distribution
differences would entail a breakdown of classical space-time making the
traditional quantum dynamics completely meaningless. The model affords a
mechanism for the evolution of such paradoxical coherent superpositions into
ensembles of pure states, each one of them corresponding then -- within a
future consistent general covariant theory -- to an unambiguous space-time.
Its features include its ability to produce an evolution of the density
matrix compatible with the expectations leading to the phenomenological
spontaneous localization models, as it was argued that they should be both
nonlinear and nonunitary \cite{bassi}. While sharing with the latter models
the non-linear non-unitary character, at variance with them, however, it
does not involve free parameters, neither it presents obstructions to its
consistent special-relativistic extension\cite{DefMaim2}. What is more,
while those models are only generically motivated by the expectation that
gravity may induce violations of the traditional unitary evolution\cite
{karolyhazy,penrose,hawking3,diosi,ghirardi,anandan,power}, we are talking
about the first consistent unified model of Newtonian gravity and
spontaneous localization. In fact it can in principle be disproved by
experiments and be used to characterize (gravitational-)decoherence-free
states\cite{DecFree}. 

Moreover, while ''there is far from universal agreement as to the meaning of
entropy -- particularly in quantum theory -- and as to the nature of the
second law of thermodynamics''\cite{wald}, the model affords, in that
respect, a very simple setting. Even in a genuinely closed system von
Neumann entropy is not constant and can then be identified, in principle,
with its thermodynamic entropy. This avoids the ill defined procedure of
coarse graining, based, as it is, on the subjective notion of macroscopic
observables\cite{vonNeumann}.

On the other hand a weak point of the model is that, in spite of being
consistent on its own, it is obtained as the Newtonian limit of a
relativistic model that, like every general covariant quantum theory, is
only poorly defined. This induces to see if it, or a well limited class of
nonrelativistic models including it, can be seen as following directly from
a set of reasonable assumptions, which is what we propose to show in the
following.

To begin with, we want to remark that the possible detection of small
deviations from the unitary dynamics of closed systems would require an
unprecedented control on the quantum state, which is likely to imply low
temperatures and energies. This suggests that the wanted modification should
first be looked for within the context of non-relativistic quantum
mechanics, where the crudest approximation corresponds to the use of
instantaneous actions at a distance. Moreover the easiest way to ensure the
consistency of a nonunitary model is by means of a unitary meta-model with
hidden degrees of freedom\cite{unruh}. In particular this allows for the
passage from the Schroedinger to the Heisenberg picture, which in general
cannot be performed for nonlinear evolutions, not even in terms of
superoperators. This is not irrelevant, as consistent (global) general
covariant extensions might be more natural in terms of the (relativistic
extension of the) Heisenberg picture, in the absence of global foliations by
space-like 3-manifolds.

As we are looking for a fundamental non-unitarity, considering a fraction of
the degrees of freedom as hidden should not be arbitrary, as in models for
environment induced decoherence. On the other hand the natural setting for
intrinsically hidden degrees of freedom, or equivalently for dynamical
algebras not being faithfully represented by the state space, and then
larger than observable algebras, is that of constrained theories. This same
standpoint can be found for instance in a recent proposal of ''a natural way
to split the Universe into two subsystems'', within quantum geometrodynamics
in extended phase space\cite{Shestakova}.

As to energy conservation, on one side fluctuations are welcome as they may
lead to traditional stationary states evolving into microcanonical mixed
states with a small spread around the initial energy value, and then to get
thermodynamical equilibrium for a closed system. On the other hand
violations of energy conservation should lead only to fluctuations, while
the hidden system should not be ''available as either a net source or a sink
of energy''\cite{unruh}, for consistency with experimental constraints\cite
{squires}.

The natural way to avoid a priory a net energy flux between observable and
hidden degrees of freedom is by means of a symmetry requirement. To be
specific the algebra of the meta-model should be the direct product of $N$
identical copies of the observable algebra, with the generator of its
meta-dynamics, the meta-Hamiltonian, symmetrical with respect to all
permutations of the copies. Of course the same symmetry should be imposed as
a constraint on meta-states.

Finally, if the interaction responsible for non-unitarity is the
gravitational one, then in its absence the meta-Hamiltonian should simply be
the sum of $N$ uncoupled ordinary gravity-free Hamiltonians $H_{0}$.
Ordinary Newton interactions should be the only means by which the $N$
copies of the observable algebra are coupled and non-unitarity comes in.

By repeating for a generic $N$ the computation performed in Refs.\cite
{defilippo1,DefMaim} for $N=2$, one sees that the most general
meta-Hamiltonian compatible with classical Newton gravity is: 
\[
H_{G}=\sum_{j=1}^{N}H_{0}[\psi _{j}^{\dagger },\psi _{j}] 
\]
\begin{eqnarray}
&&-\frac{m^{2}G}{2N}\sum_{j\neq k}^{N}\int dxdy\frac{\psi _{j}^{\dagger
}(x)\psi _{j}(x)\psi _{k}^{\dagger }(y)\psi _{k}(y)}{|x-y|}\left( 1+\frac{%
\varepsilon }{N-1}\right)  \nonumber \\
&&-\frac{m^{2}G}{2N}\sum_{j=1}^{N}\int dxdy\frac{\psi _{j}^{\dagger }(x)\psi
_{j}(x)\psi _{j}^{\dagger }(y)\psi _{j}(y)}{|x-y|}\left( 1-\varepsilon
\right) .  \label{metahamiltonian}
\end{eqnarray}
Here $G$ and $m$ denote the gravitational constant and the mass, for
simplicity of a single particle species, while $\psi _{1}$, $\psi _{2}$,...,$%
\psi _{N}$ are $N$ commuting copies of the one particle annihilation
operator and each single product $\psi _{j}^{\dagger }(x)\psi _{j}(x)$ is a
shorthand for a scalar combination of spin components.

As physical states are obtained from meta-states by tracing out the hidden
degrees of freedom, say for $j=2,3,...,N$, pure and mixed states
respectively correspond to product and entangled meta-states. If $\left|
0\right\rangle _{j}$ denotes the vacuum of $\psi _{j}$ and $F[\psi
_{j}^{\dagger }]$ is a homogeneous operator corresponding to the creation of
a given number of $j$ meta-particles, the pure state represented by 
\[
F[\psi _{1}^{\dagger }]\left| 0\right\rangle _{1} 
\]
corresponds to the product meta-state 
\[
F[\psi _{1}^{\dagger }]\left| 0\right\rangle _{1}\otimes F[\psi
_{2}^{\dagger }]\left| 0\right\rangle _{2}\otimes ...\otimes F[\psi
_{N}^{\dagger }]\left| 0\right\rangle _{N}\text{.} 
\]

If the meta-Hamiltonian were just the first sum of $N$ uncoupled terms in
Eq. (\ref{metahamiltonian}), this meta-state would stay separable. Newtonian
interactions, for $\varepsilon \neq 1-N$, give rise instead to entangled
meta-states, which belong to a proper subspace $S$ of the product of the $N$
Fock spaces. The meta-state space $S$ is obtained from the meta-vacuum $%
\left| \left| 0\right\rangle \right\rangle \equiv \left| 0\right\rangle
_{1}\otimes \left| 0\right\rangle _{2}\otimes ...\otimes \left|
0\right\rangle _{N}$ by applying operators constructed in terms of the
products \ $\psi _{1}^{\dagger }(x_{1})\psi _{2}^{\dagger }(x_{2})...\psi
_{N}^{\dagger }(x_{N})$ and symmetrical with respect to the permutations of $%
\psi _{1}^{\dagger }$, $\psi _{2}^{\dagger }$, ..., $\psi _{N}^{\dagger }$.
If $P$ denotes the generic permutation, a generic element $\left| \left|
\Phi \right\rangle \right\rangle $ of $S$ fulfills the constraints 
\begin{equation}
\int dx\left( \psi _{j}^{\dagger }(x)\psi _{j}(x)-\psi _{k}^{\dagger
}(x)\psi _{k}(x)\right) \left| \left| \Phi \right\rangle \right\rangle
=\left( P-\widehat{{\bf 1}}\right) \left| \left| \Phi \right\rangle
\right\rangle ={\bf 0},\;\;j,k=1,2,...,N\text{.}
\end{equation}

Of course, like within any constrained theory, the constrained meta-state
space $S$ gives an unfaithful representation of the original dynamical
algebra. In particular, as the constrained meta-state space does not
distinguish between $F[\psi _{1},\psi _{1}^{\dagger }]$ and $F[\psi
_{j},\psi _{j}^{\dagger }]$, the $N-1$ copies of the observable algebra for $%
j=2,...,N$ are consistently referred to hidden degrees of freedom\cite{unruh}%
. While $\varepsilon =0$ in Ref.\cite{DefMaim}, a simple and appealing
alternative (avoiding singularities in the gravitational collapse)\cite
{DefMaim} is $\varepsilon =1$\cite{defilippo1}, where the Newton interaction
is purely of the nonunitary type. Some of the physical properties for $N=2$
and $\varepsilon =0,1$ were analyzed in Refs. \cite
{defilippo1,DefMaim,defilippo2,DefMaim1,DefMaimRob}, with qualitatively
equivalent results in the two cases. The limit $N\rightarrow \infty $ was
shown to be equivalent to the Schroedinger-Newton model \cite{defilippo4}.
Finally only for $\varepsilon >1-N$ the model properties are qualitatively
of the same kind as for $\varepsilon =0,1$.

Even though this ends the characterization of the model, we want to show how
the case $N=2$ can be inferred, by means of some simple hypotheses, just on
thermodynamical grounds. First observe that at the equilibrium the (inverse)
temperatures of the physical and the hidden systems should coincide: 
\begin{equation}
\frac{\partial S_{p}}{\partial E_{p}}=\frac{\partial S_{h}}{\partial E_{h}},
\end{equation}
where $E_{p}$, $E_{h}$ respectively denote the energy of the physical and
the hidden system at equilibrium, while $S_{p},$ $S_{h}$ denote their
thermodynamic entropies. If the physical thermodynamic entropy is identified
with the entanglement entropy with the hidden degrees of freedom, it is only
natural to assume the same for the entropy of the hidden system, by which $%
S_{p}\left( E_{p}\right) =S_{h}\left( E_{h}\left( E_{p}\right) \right) $, as
it happens for the entanglement entropy of every bipartite system. Then 
\begin{equation}
\frac{\partial S_{p}}{\partial E_{p}}=\frac{\partial S_{h}}{\partial E_{h}}%
\frac{\partial E_{h}}{\partial E_{p}},
\end{equation}
which, together with the equilibrium condition and by a coherent choice of
the physical and hidden energy zero, in its turn implies $E_{h}=E_{p}\equiv E
$, by which $S_{p}(E)=S_{h}(E)$, namely the physical and the hidden system
have one and the same thermodynamics. It is only natural then to assume that
they are dynamically equivalent too. This of course is just one step from
getting Eq.(\ref{metahamiltonian}) for $N=2$. As to the case of an arbitrary 
$N$, it corresponds to assuming that the entropy of the hidden system is the
sum of the entropies of $N-1$ weakly interacting subsystems, each one of
them being an entanglement entropy.

While the model is nonlinear -- the product meta-state corresponding to a
linear combination of pure states is different from the entangled linear
combination of the meta-states corresponding to them separately -- like the
phenomenological localization models proposed so far, at variance with them
it is non-Markovian, in accordance with the suggestion of Ref. \cite{unruh}.
If we perform a traditional complete set of measurements on a given system,
then we know both its (pure) physical state and the corresponding product
meta-state. That is enough to determine its future evolution as a closed
system, which leads to mixed physical states. However, the mere knowledge of
such a physical mixed state at a given instant of time would not allow to
know the meta-state and then the future evolution. This means that we have
an evolution with memory: the system ''remembers'' the pure state it is
evolving from. It is worthwhile remarking that, when we say
''non-Markovian'', we are implying that the evolution of the system (mixed)
state has a hereditary character and then that any possible unraveling would
correspond to a non-Markovian quantum stochastic process. Actually the
situation is very peculiar. In fact, if we consider other physical
hereditary systems, like Feynman-Wheeler electrodynamics (where not even the
knowledge of the past is enough), or more simply (apart from the well known
consistency problems) electrodynamics with retarded potentials, the
situation is quite different: the knowledge of the system state at a given
instant is never enough in order to reconstruct the system future.

To be more specific about the hereditary character of the model, consider an
initial pure state represented by the normalized ket $\left| \varphi
_{1}\right\rangle $. Assume that the corresponding product meta-ket $\left|
\varphi _{1}\right\rangle \left| \varphi _{1}\right\rangle $, for $N=2$,
evolves, for simplicity, into the entangled meta-state represented by 
\begin{equation}
\alpha \left| \chi \right\rangle \left| \chi \right\rangle +\beta \left|
\omega \right\rangle \left| \omega \right\rangle ,\;\;\left\langle \chi \mid
\chi \right\rangle =\left\langle \omega \mid \omega \right\rangle
=1,\;\left\langle \omega \mid \chi \right\rangle =0,\;\alpha ,\beta \in
R\;\alpha ^{2}+\beta ^{2}=1,\;\;  \label{entangled}
\end{equation}
where first and second factors within tensor products refer respectively to
the physical and the hidden Hilbert space. If $\left| \overline{\chi }%
\right\rangle $ and $\left| \overline{\omega }\right\rangle $ represent the
states obtained respectively from $\left| \chi \right\rangle $ and $\left|
\omega \right\rangle $ by time reversal, then the time evolution of the
entangled meta-state $\alpha \left| \overline{\chi }\right\rangle \left| 
\overline{\chi }\right\rangle +\beta \left| \overline{\omega }\right\rangle
\left| \overline{\omega }\right\rangle $ gives rise to the product metastate 
$\left| \overline{\varphi _{1}}\right\rangle \left| \overline{\varphi _{1}}%
\right\rangle $, where $\left| \overline{\varphi _{1}}\right\rangle $
represents the time reversed initial pure state. It is immediate to prove
that this implies that the time evolution of $\left| \overline{\chi }%
\right\rangle \left| \overline{\chi }\right\rangle $ and $\left| \overline{%
\omega }\right\rangle \left| \overline{\omega }\right\rangle $ can be
represented as 
\begin{eqnarray}
\left| \overline{\chi }\right\rangle \left| \overline{\chi }\right\rangle
&\longrightarrow &\sum_{i,j}c_{i,j}\left| \overline{\varphi _{i}}%
\right\rangle \left| \overline{\varphi _{j}}\right\rangle  \nonumber \\
\left| \overline{\omega }\right\rangle \left| \overline{\omega }%
\right\rangle &\longrightarrow &\sum_{i,j}\frac{\delta _{i,1}\delta
_{j,1}-\alpha c_{i,j}}{\beta }\left| \overline{\varphi _{i}}\right\rangle
\left| \overline{\varphi _{j}}\right\rangle
\end{eqnarray}
with $\left\langle \overline{\varphi _{i}}\mid \overline{\varphi _{j}}%
\right\rangle =\delta _{i,j}$. On the other hand the metastates represented
by $e^{i\theta }\alpha \left| \overline{\chi }\right\rangle \left| \overline{%
\chi }\right\rangle +\beta \left| \overline{\omega }\right\rangle \left| 
\overline{\omega }\right\rangle $, where $\theta $ is an arbitrary phase,
all correspond to one and the same state, $\alpha ^{2}\left| \overline{\chi }%
\right\rangle \left\langle \overline{\chi }\right| +\beta ^{2}\left| 
\overline{\omega }\right\rangle \left\langle \overline{\omega }\right| $.
However, for the linearity of the meta-dynamics, \ such states evolve into
the metastates 
\begin{equation}
\left| \left| \Psi \left( \theta \right) \right\rangle \right\rangle
=\sum_{i,j}\left[ \alpha c_{i,j}\left( e^{i\theta }-1\right) +\delta
_{i,1}\delta _{j,1}\right] \left| \overline{\varphi _{i}}\right\rangle
\left| \overline{\varphi _{j}}\right\rangle ,
\end{equation}
corresponding, by tracing out from $\left| \left| \Psi \left( \theta \right)
\right\rangle \right\rangle \left\langle \left\langle \Psi \left( \theta
\right) \right| \right| $ the hidden factor, to the generally mixed states 
\begin{equation}
Tr_{2}\left| \left| \Psi \left( \theta \right) \right\rangle \right\rangle
\left\langle \left\langle \Psi \left( \theta \right) \right| \right|
=\sum_{i,j,k}\left[ \alpha c_{i,j}\left( e^{i\theta }-1\right) +\delta
_{i,1}\delta _{j,1}\right] \left[ \alpha c_{k,j}^{\ast }\left( e^{-i\theta
}-1\right) +\delta _{k,1}\delta _{j,1}\right] \left| \overline{\varphi _{i}}%
\right\rangle \left\langle \overline{\varphi _{k}}\right|
\end{equation}
This state coincides with the pure state $\left| \overline{\varphi _{1}}%
\right\rangle \left\langle \overline{\varphi _{1}}\right| $ if and only if $%
\theta =0$, which proves that the knowledge of a (mixed) state is not enough
to determine its evolution in time. Parenthetically, if the simplifying
assumption in Eq. (\ref{entangled}) of a linear combination of only two
product states is replaced by a more realistic infinite sum, the phase $%
\theta $ is replaced by infinitely many phases. This illustrates the
rationale for a very low probability of processes with an entropy decrease,
in particular to a vanishing value. In fact, among the infinitely many
possible choices, only for all vanishing phases the analogue of the state $%
\alpha ^{2}\left| \overline{\chi }\right\rangle \left\langle \overline{\chi }%
\right| +\beta ^{2}\left| \overline{\omega }\right\rangle \left\langle 
\overline{\omega }\right| $ would evolve into the pure state $\left| 
\overline{\varphi _{1}}\right\rangle \left\langle \overline{\varphi _{1}}%
\right| $, this requiring a very unlikely fine tuning.

Finally, to clarify the relevance of the present model to the quantum
foundations of thermodynamics, consider a closed system, for simplicity
small enough to make gravity irrelevant in the usual setting of a unitary
dynamics, but large enough to be governed by extensive thermodynamics. An
eigenstate of the physical Hamiltonian evolves into a mixed state, which in
the absence of ergodicity-breaking symmetries we expect to be equivalent, at
a generic instant of time far enough in the future, to a microcanonical
ensemble with an energy width $\Delta E$, apart from thermodynamic
fluctuations. The strength of the thermalizing interaction -- here the
gravitational constant $G$\ -- determines only the thermodynamically
irrelevant width $\Delta E$.

One might object that in the usual kinetic theory gravity plays no role and,
in spite of that, one gets sensible results as to thermalization times. The
objection would be the analogue of the claim that environment induced
decoherence solves the measurement problem, contrary to the opinion even of
a founding father like Zeh\cite{zeh}. There too the apparent nonunitarity
makes the role of any fundamental nonunitarity marginal for ordinary
situations, where the former has an overwhelming influence. This is the more
true for our model, where the presence of a mass threshold forces to
consider macroscopic systems to look for a relevant fundamental nonunitarity.

While we intended to get rid of the subjective elements in the quantum
foundations of thermodynamics, one should remember that even for closed
systems only local quantities are usually accessible. Consider for instance
the evolution of a pure product state of a gas system, where one factor
refers to a subset of molecules, forgetting for simplicity
indistinguishability. Intermolecular interactions with the other molecules
soon produce an entangled state, which, as far as measurements are performed
on that subset only, corresponds to a mixed state. In particular this allows
to define a subjective entropy as a sum of entropies of constituent
subsystems, even though the state of the complete system stays pure.

The emerging picture consists then in a subjective entropy growth followed
by a slower fundamental one. It should be remarked that ordinary
interactions are effective in giving rise not only to a subjective faster
kinetics, but also to a faster growth of von Neumann entropy. In fact they
produce entanglement between the degrees of freedom referring to different
space scales, whereas only larger space scales are mainly involved in the
gravitational interaction with the hidden degrees of freedom.

\end{document}